\documentclass[12pt]{article}
\usepackage{epsfig}
\begin{document}
\begin{titlepage}
\begin{center}


\vspace{2cm}

{\Large \bf On the $\pi\pi$ contribution to the QCD sum rules for
the light tetraquark} \\
\vspace{0.50cm}
\renewcommand{\thefootnote}{\fnsymbol{footnote}}
Hee-Jung Lee$^{a}$ \footnote{hjl@chungbuk.ac.kr},
N.I. Kochelev$^{b}$\footnote{kochelev@theor.jinr.ru}

\vspace{0.50cm}

{(a) {\it School of Science Education, Chungbuk National
University,\\
Cheongju Chungbuk 361-763, South Korea}} \vskip 1ex

{(b) {\it Bogoliubov Laboratory of Theoretical Physics,\\
Joint Institute for Nuclear Research, Dubna, Moscow region, 141980 Russia}}

\end{center}
\vskip 0.5cm

\centerline{\bf Abstract} We perform a QCD sum rule analysis for the
$f_0$ light tetraquark taking into account the contribution arising
from the two pion intermediate state. With the interpolating
currents of the different chiral combinations of scalar and
pseudoscalar diquarks, it is demonstrated that the interpolating
current with maximum chirality has a large coupling to the two pion
state, but the current with zero chirality interacts only weekly
with this state. Taking into account the form factor in the
$f_0$--two pion vertex, it is shown that the $f_0$--coupling to the
two pion state leads to an increase of the lightest tetraquark mass
by a value of about 100 MeV. The analysis of the resulting sum rule
shows that the $\sigma(f_0(600))$--meson state might be treated as
the four--quark bound state in the instanton field which has a
rather strong coupling to the two pion state.

\vskip
0.3cm \leftline{Pacs: 11.55.Hx, 12.38.Aw, 12.38.Lg, 14.40-n}
\leftline{Keywords: QCD sum rule, scalar meson, tetraquark,
instanton}

\vspace{1cm}

\end{titlepage}

\setcounter{footnote}{0}
\section{Introduction}

One of the direct ways to investigate the properties of the exotic
states is the QCD sum rule (SR) approach. Recently, this method was
applied to the study of the scalar tetraquark states with different
interpolating currents. It was shown that the light $0^{++}$ mesons
($\sigma (f_0(600))$, $\kappa(800)$, $a_0(980)$ and $f_0(980)$)
might be interpreted as four--quark exotic states
\cite{Brito:2004tv,Wang:2005cn,Chen:2007zzg,Chen:2006hy,leekoch}.
The investigations give additional support to possible large not
$q$--$\bar q$ component in these states in the line of their
previous study within different constituent quark models (see
\cite{models} and references therein) and the approach based on
the $1/N_c$ expansion \cite{pelaez}. In most of these SR analysis only
the exotic resonance contribution to the phenomenological part of the SR has
been considered within the narrow width approximation for the pole.
However, it is well known that the tetraquark can couple strongly to
two meson colorless states. Such coupling is super--allowed
according to the OZI rule and might be responsible for the large
observed width of the lightest candidate for tetraquark the
$\sigma(f_0(600))$ meson. Therefore one may expect a rather large
effect of such coupling on the extracted properties of the
tetraquark within both QCD SR and lattice approaches
\cite{lattice1}. We should emphasize that the problem of the
possible large contribution of intermediate hadronic states to the
correlator of the multiquark current is quite general and, for
example, has been discussed recently for the pentaquark case in
ref.~\cite{suhong}.

When the QCD sum rule is applied to the tetraquarks, it is well
known that the operators of higher dimensions could yield large
contributions in the operator product expansion (OPE) and would
spoil convergence in the OPE~\cite{Lee:2005hs,Matheus07}. In
our previous paper \cite{leekoch}, it was demonstrated that the
interpolating currents with equal weights of scalar and pseudoscalar
diquarks yield strong cancelation of the contributions coming from
high dimension operators and direct instantons. It was shown that
such cancelation is related to the specific chirality structure of
the interpolating current. As a result,  we can avoid the problem of
huge contribution from high dimension operators to the tetraquark
SR. In this way, a quite stable SR for the light tetraquark  meson
with $u\bar u d\bar d$ quark content has been obtained.

In this here we extend the previous study to include the two pion
intermediate contribution to the QCD SR. In section II the
contribution to the SR  from the two pion state for the
interpolating current with arbitrary mixing of scalar and
pseudoscalar $ud$-diquark will be obtained  by using soft pion
theorems. In section III the numerical analysis of the corresponding
SR will be done analyzing the effect of the form factor in the
tetraquark--two pion vertex, and in section IV we will give the
summary of our results.

\section{Two pion contribution to QCD sum rule for light
tetraquark}

The starting point of the QCD sum rule for the scalar meson is the
dispersion relation of the correlator:
\begin{equation}
\Pi(q^2)=\frac{1}{\pi}\int_0^\infty ds^2\ \frac{{\rm
Im}\Pi(s^2)}{s^2-q^2}\ , \label{corr}
\end{equation}
where the correlator is defined by
\begin{equation}
\Pi(q^2)=i\int d^4x\ e^{iq\cdot x}\langle
0|TJ_{f_0}(x)J_{f_0}^{\dagger}(0)|0\rangle
\end{equation}
where $J_{f_0}(x)$ is the interpolating current for the scalar
meson. The imaginary part of the correlator is given by the spectral
sum over various intermediate particle states
\begin{eqnarray}
\frac{1}{\pi}\ {\rm Im}\Pi(q^2)&=&(2\pi)^3\sum_n
\delta^4(q-P_n)\langle 0|J_{f_0}(0)|n\rangle \langle n
|J_{f_0}^\dagger(0)|0 \rangle \ .
\end{eqnarray}
In the single narrow width resonance approximation and with the
assumption of hadron--quark duality, the contribution to
phenomenological part of the SR coming from the first three diagrams
pictured in Fig. 1 is usually considered. In this case the imaginary
part is the following,
\begin{equation}
\frac{1}{\pi}{\rm Im}\Pi(s^2)=2f_{f_0}^2 m_{f_0}^8\delta(s^2-m_{f_0}^2)
+\theta(s^2-s_0^2)\frac{1}{\pi}{\rm Im}\Pi^{OPE}(s^2), \label{Im}
\end{equation}
where $f_{f_0}$ is the residue of the resonance, $m_{f_0}$ is its
mass, $s_0$ is the continuum threshold and $\Pi^{OPE}$ is the
correlator within the standard operator expansion (OPE).

\begin{figure}[h]
\centerline{\epsfig{file=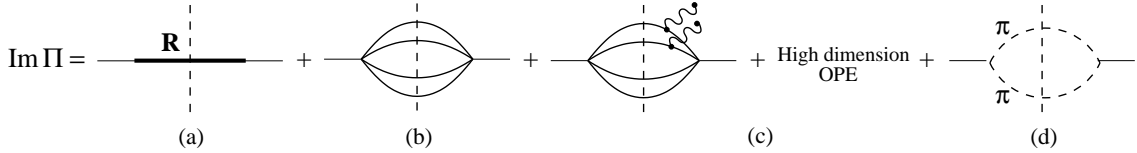,width=15cm,angle=0}} \caption{The
contributions to the phenomenological part of the SR for a light
tetraquark: (a) resonance, (b) and (c) continuum, and (d) two pion
contributions.}
\end{figure}

In ref.~\cite{leekoch} we have shown that, the correlator of
the interpolating current consisting of scalar and pseudoscalar
diquarks
\begin{equation}
J_{f_0}=\alpha J_S+\beta J_{PS},  \label{current}
\end{equation}
where
\begin{eqnarray}
J_S&=&\epsilon_{abc}\epsilon_{ade}
(u^T_b\Gamma_{S}d_{c})(\bar{u}_d\overline\Gamma_{S}\bar{d}^T_{e})\ ,
\nonumber\\
J_{PS}&=&\epsilon_{abc}\epsilon_{ade}
(u^T_b\Gamma_{PS}d_{c})(\bar{u}_d\overline\Gamma_{PS}\bar{d}^T_{e})\ ,
\end{eqnarray}
and $\Gamma_S=C\gamma^5,\ \Gamma_{PS}=C,\ $
$\overline{\Gamma}_i=\gamma^0\Gamma^\dagger_i\gamma^0$, has very
specific properties for a particular choice of the mixing parameters
$\alpha=\pm \beta$. Indeed, it was found that these choice of mixing
parameters leads to a cancelation of the contribution from the high
dimensions operators in the OPE, as well as that of some dangerous
direct instanton contributions.

The contribution from the two pion state to the imaginary part of
the correlator (the last diagram in Fig.1) is presented by phase space
product of the two pions as
\footnote{We do not consider the
possible effects of the interaction between two pions for simplicity.
This assumption  was also used in the paper \cite{Dosch:2002rh} to
estimate the effect in the SR arising from the coupling of two
uncorrelated pions to the quark--antiquark $\sigma$--meson state.}
\begin{eqnarray}
\frac{1}{\pi}\ {\rm Im}\Pi^{2\pi}(q^2)&=&(2\pi)^3|\lambda_{2\pi}|^2
\int\frac{d^3p_1}{(2\pi)^3 2E_1}\int \frac{d^3p_2}{(2\pi)^3 2E_2}\
\delta^4(q-p_1-p_2)
\nonumber\\
&=&\frac{|\lambda_{2\pi}|^2}{16\pi^2}\sqrt{1-\frac{4m_\pi^2}{q^2}}\
\theta(q^2-4m_\pi^2),
\end{eqnarray}
where $\lambda_{2\pi}=\langle 0 |J_{f_0}(0)|\pi\pi\rangle$ is
the correspondent residue.

To obtain the value of $\lambda_{2\pi}$, we rewrite the
interpolating current in terms of pion fields by using the Fierz
transformation with the two flavor quark field $\psi$
\footnote{ A similar approach was used in ref.~\cite{KV}
to obtain the instanton contribution to the weak decay
$K\rightarrow \pi\pi$.}
\begin{eqnarray}
J_{f_0}&=&\alpha J_S+\beta J_{PS}
\nonumber\\
&=&\frac{1}{16}\bigg[(\alpha-\beta)\bigg((\bar{\psi}\psi)^2
-(\bar{\psi}\vec{\tau}\psi)^2
+(\bar{\psi}\gamma_5\psi)^2-(\bar{\psi}\gamma_5\vec{\tau}\psi)^2\bigg)
\nonumber\\
&& +(\alpha+\beta)\bigg((\bar{\psi}\gamma^5\gamma_\rho\psi)^2
-(\bar{\psi}\gamma^5\gamma_\rho\vec{\tau}\psi)^2
+(\bar{\psi}\gamma_\rho\psi)^2
-(\bar{\psi}\gamma_\rho\vec{\tau}\psi)^2\bigg)
\nonumber\\
&&+\frac{1}{2}(\alpha-\beta)
\bigg((\bar{\psi}\sigma_{\rho\sigma}\psi)^2-
(\bar{\psi}\sigma_{\rho\sigma}\vec{\tau}\psi)^2\bigg)\bigg]\ .
\end{eqnarray}
The relevant part in the Eq.8 which
includes the pionic interpolating currents is
\begin{eqnarray}
&&J_{f_0}^\pi=-\frac{1}{16}\bigg[(\alpha-\beta)\bigg(
(\bar{u}\gamma^5d+\bar{d}\gamma^5u)^2-(\bar{u}\gamma^5d-\bar{d}\gamma^5u)^2
+(\bar{u}\gamma^5u-\bar{d}\gamma^5d)^2\bigg)
\nonumber\\
&&+(\alpha+\beta)\bigg( (\bar{u}\gamma^5\gamma_\mu
d+\bar{d}\gamma^5\gamma_\mu u)^2 -(\bar{u}\gamma^5\gamma_\mu
d-\bar{d}\gamma^5\gamma_\mu u)^2 +(\bar{u}\gamma^5\gamma_\mu
u-\bar{d}\gamma^5\gamma_\mu d)^2\bigg)\bigg]\ .
\end{eqnarray}

By transforming well--known PCAC relations at
the limit $m_u=m_d=m_q$,
\begin{eqnarray}
\partial^\mu A^1_\mu=im_q(\bar{u}\gamma^5d+\bar{d}\gamma^5u),\ \
\partial^\mu A^2_\mu=m_q(\bar{u}\gamma^5d-\bar{d}\gamma^5u), \ \
\partial^\mu
A^3_\mu=im_q(\bar{u}\gamma^5u-\bar{d}\gamma^5d)\nonumber
\end{eqnarray}
into operator forms, $\partial^\mu A_\mu^a=f_\pi m_\pi^2 \phi^a$ and
$A_\mu^a=ip_\mu^a f_\pi\phi^a$ with the pion field of isospin index
 $a$, $\phi^a$,
we can rewrite the above interpolating current
in terms of pion fields as
\begin{eqnarray}
J_{f_0}^\pi&=&\frac{1}{16}\bigg[\frac{f_\pi^2
m_\pi^4}{m_q^2}(\alpha-\beta)
\bigg(2\phi_{\pi^+}\phi_{\pi^-}+\phi_{\phi^0}^2\bigg)
\nonumber\\
&&+4(\alpha+\beta)f_\pi^2\bigg(2p_{\pi^+}\cdot
p_{\pi^-}\phi_{\pi^+}\phi_{\pi^-} +p^1_{\pi^0}\cdot
p^2_{\pi^0}\phi_{\pi^0}\phi_{\pi^0}\bigg)\bigg]\nonumber.
\end{eqnarray}
Therefore, we have
\begin{equation}
|\lambda_{2\pi}|^2=6\bigg[(\alpha-\beta)^2
\bigg(\frac{\langle\bar{q}q\rangle^2}{4f_\pi^2}\bigg)^2
+(\alpha+\beta)^2\bigg(\frac{f_\pi^2}{4}\bigg)^2\frac{(q^2-2m_\pi^2)^2}{4}\bigg]
\end{equation}
where the Gell--Mann--Oakes--Renner relation
\begin{equation}
f_\pi^2m_\pi^2=-2m_q\langle\bar{q}q\rangle \nonumber
\end{equation}
has been used. The imaginary part of the correlator
coming from the two intermediate pion state becomes
\begin{eqnarray}
\frac{1}{\pi}\ {\rm Im}\Pi^{2\pi}(q^2)
&=&\frac{6}{16^2\pi^2}\bigg[
\frac{\langle\bar{q}q\rangle^4}{f_\pi^4}(\alpha-\beta)^2
+\frac{f_\pi^4}{4}(q^2-2m_\pi^2)^2(\alpha+\beta)^2\bigg]
\nonumber\\
&&\times\sqrt{1-\frac{4m_\pi^2}{q^2}}\ \theta(q^2-4m_\pi^2).
\label{im2pi}
\end{eqnarray}

Incorporating this contribution from the two pion state,
the QCD sum rule for the light tetraquark becomes
\begin{eqnarray}
&&\frac{1}{\pi}\int_{0}^{s_0^2}ds^2\
e^{-s^2/M^2}{\rm Im}\Pi^{OPE}(s^2)+\hat{B}[\Pi^{I+\bar{I}}(q)]
-\frac{1}{\pi}\int_{4m_\pi^2}^{s_0^2}ds^2\ e^{-s^2/M^2}{\rm Im}\Pi^{2\pi}(s^2)
\nonumber\\
&&=2f_{f_0}^2m_{f_0}^8e^{-m_{f_0}^2/M^2},
\label{SR}
\end{eqnarray}
where $\hat{B}[\Pi^{I+\bar{I}}(q^2)]$ means the Borel transformed instanton effect.
Here the contributions from the two pion state and the continuum are transferred
to the OPE side in the sum rule.

In the numerical analysis below we will use the results for the
standard OPE and direct instanton contributions to the correlator
obtained in our previous paper \cite{leekoch}. For massless $u$--
and $d$--quarks the relevant parts of the correlator are given by
\begin{eqnarray}
\frac{1}{\pi}\ {\rm Im}\
\Pi^{OPE}(q^2)&=&(\alpha^2+\beta^2)\bigg[\frac{(q^2)^4}{2^{12}\cdot5\cdot3\pi^6}\
+\frac{\langle g^2G^2\rangle}{2^{11}\cdot3\pi^6}(q^2)^2\bigg]
\nonumber\\
&&+(\alpha^2-\beta^2)\bigg[\frac{\langle\bar{q}q\rangle^2}{12\pi^2}\
q^2-\frac{\langle\bar{q}q\rangle \langle
ig\bar{q}\sigma\cdot Gq\rangle}{12\pi^2}
\nonumber\\
&&+\frac{59(\langle ig \bar{q}\sigma\cdot
Gq\rangle)^2}{2^{9}\cdot3^2\pi^2}\ \delta(q^2)
+\frac{7\langle g^2
G^2\rangle\langle\bar{q}q\rangle^2}{2^5\cdot3^3\pi^2}\ \delta(q^2)\bigg]\  ,
\label{ImP}
\end{eqnarray}
and
\begin{eqnarray}
\Pi^{I+\bar{I}}(q)&=&(\alpha^2-\beta^2)\frac{32n_{eff}\rho_c^4}
{\pi^8m_q^{*2}}f_6(q)
\nonumber\\
&&+[19(\alpha^2+\beta^2)-6\alpha\beta]
\frac{n_{eff}\rho_c^4\langle\bar{q}q\rangle^2}{18\pi^4m_q^{*2}}
f_0(q),
\label{instSR}
\end{eqnarray}
where $n_{eff}$ is the effective instanton density, $m_q^*$ is the mass
parameter in the quark zero mode Green's function in the instanton field, and
$\rho_c$ is average instanton size. The functions $f_6(q),\ f_0(q)$
are defined by
\begin{eqnarray}
f_6(q)&=&\int d^4z_0 \int d^4x \frac{e^{iq\cdot x}}
{x^6[z_0^2+\rho_c^2]^3[(x-z_0)^2+\rho_c^2]^3}\ ,
\nonumber\\
f_0(q)&=&\int d^4z_0 \int d^4x \frac{e^{iq\cdot x}}
{[z_0^2+\rho_c^2]^3[(x-z_0)^2+\rho_c^2]^3}\ . \label{func}
\end{eqnarray}
There are two types of singularities in these functions.
One arises from the origin and another from a finite distance from
the origin. Note that we subtract the contribution from the pole at the origin
in order to avoid double counting with the contributions from the condensates
in the standard OPE~\cite{Chibisov:1996wf,hjl2006}.

It is evident from
Eq.\ref{ImP} and Eq.\ref{instSR} that if the
relation $\alpha^2=\beta^2$ holds true, then most part of the high
dimension operators in OPE and part of direct instanton contribution
 disappear from the SR. In this case only the perturbative and
gluon condensate chirality conserving contributions remain
 in the OPE due to the specific chirality structure of the tetraquark
interpolating current Eq.~\ref{current}. Indeed, this current can be
decomposed into two parts with different chirality structures
\cite{leekoch}
\begin{eqnarray}
J_{f_0}&\sim&-(\alpha-\beta)(u_L^TCd_L\bar{u}_LC\bar{d}_L^T
+u_R^TCd_R\bar{u}_RC\bar{d}_R^T)
\nonumber\\
&&+(\alpha+\beta)(u_R^TCd_R\bar{u}_LC\bar{d}_L^T+u_L^TCd_L\bar{u}_RC\bar{d}_R^T).\
\end{eqnarray}

For the values
$\alpha=\beta$ or $\alpha=-\beta$ the current carries zero
value of  chirality or four units of the chirality, respectively.
We call the first(second) case  as the minimum(maximum) chirality
current.
It is easy to verify that for these particular cases, most of the
high dimensional condensates and direct instanton contributions are
forbidden. As result, one has a good convergence
of the OPE  and the possible
stability of the SR (see discussion in \cite{leekoch}). Therefore,
below we consider the two pion contribution to the SR only for these
values of the mixing parameters.

\section{Numerical analysis of the sum rule with two pion contribution}
For the numerical analysis of the SR for $\alpha^2=\beta^2$ we use
the following value for the gluon condensate and the average size of
the instanton
\begin{eqnarray}
\langle g^2G^2\rangle=0.5\ {\rm GeV}^4, \ \
\rho_c=1.6\ {\rm GeV}^{-1}\ ,
\end{eqnarray}
and the relation between instanton parameters given by the simplest
version of Shuryak's instanton liquid model \cite{shuryak98,dorokhov}
\begin{equation}
\frac{2n_{eff}}{m_q^{*2}}=\frac{3}{2\pi^2\rho_c^2}\ .
\end{equation}
\begin{figure}[htb]
\begin{minipage}[c]{8cm}
\hspace*{0.5cm} \psfig{file=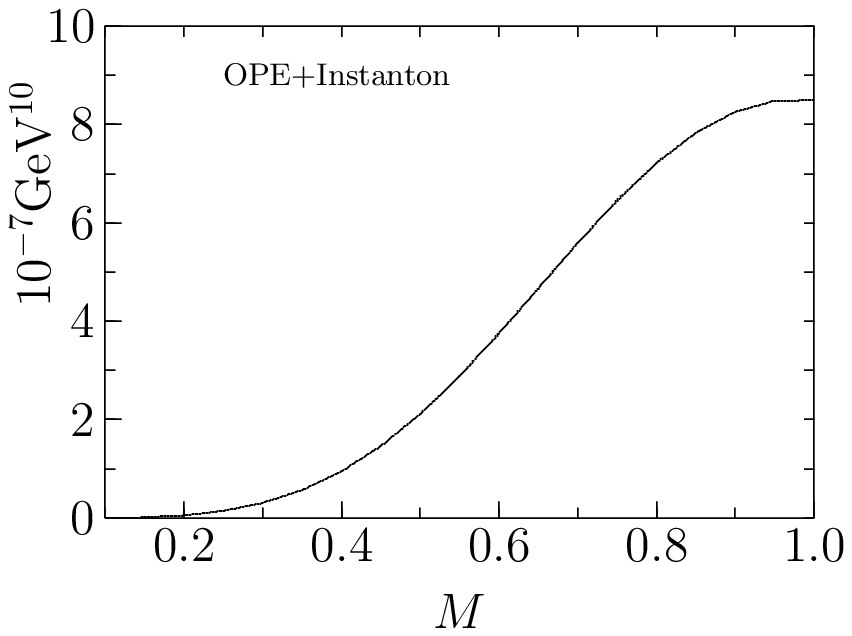,width=6cm,height=5cm}
\caption{The OPE and direct instanton contribution to the SR
with $\alpha=-\beta=1 $.}
\end{minipage}
\hspace*{0.5cm}
\begin{minipage}[c]{8cm}
\vskip 1.0cm
\hspace*{0.5cm}
\psfig{file=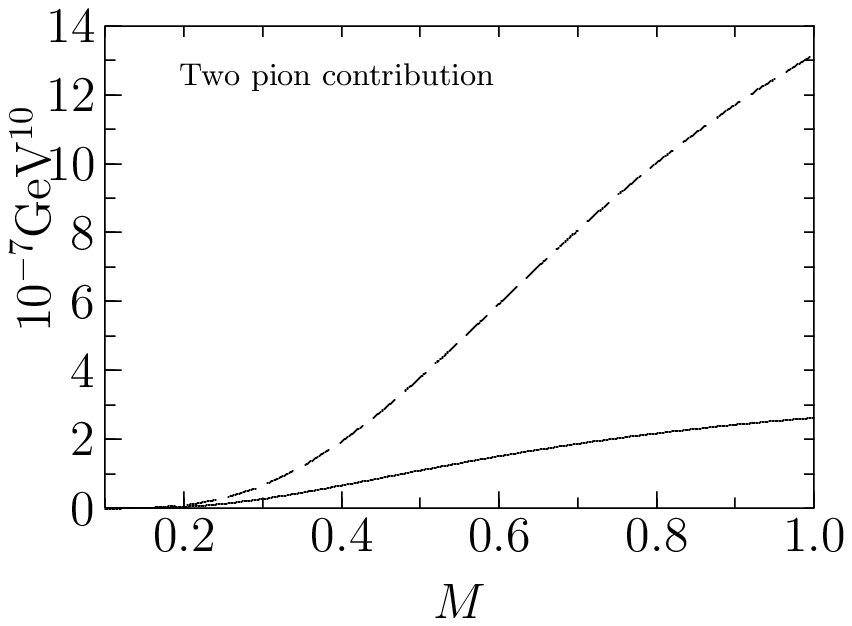,width=6cm,height=5cm}
\caption{The two pion contribution to the phenomenological side of
the SR with $\alpha=-\beta=1 $. The solid (dashed) line
contains(does not contain) the effect of the form factor.}
\end{minipage}
\end{figure}
We fix the value of the threshold by $s_0=1$ GeV since our results
below show only weak dependence on this parameter within the
interval $s_0=1.\div 1.5$ GeV.
\begin{figure}[h]
\begin{minipage}[c]{8cm}
\hspace*{0.5cm}
\psfig{file=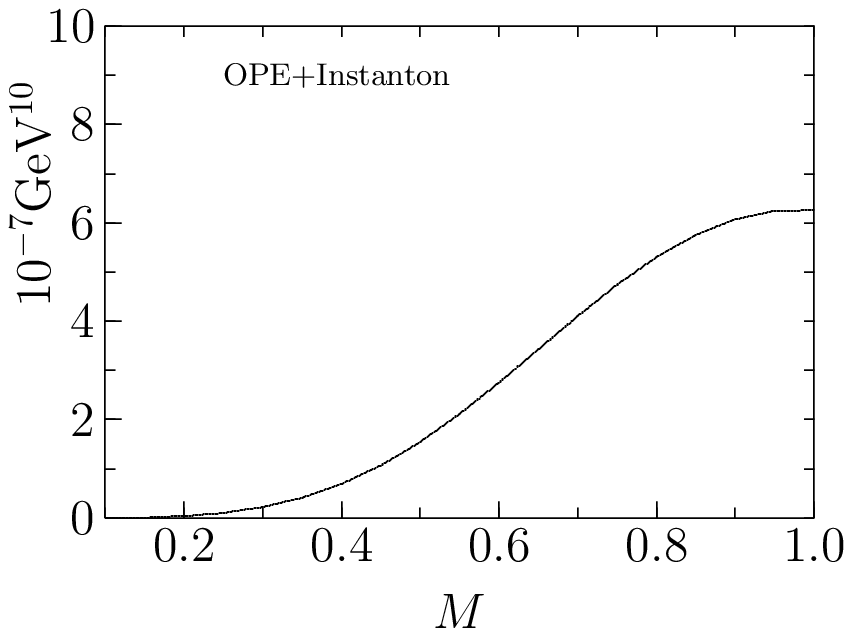,width=6cm,height=5cm}
\caption{The OPE and direct instanton contribution to the SR with
$\alpha=\beta=1 $.}
\end{minipage}
\hspace*{0.5cm}
\begin{minipage}[c]{8cm}
\vskip 1.5cm
\hspace*{0.5cm} \psfig{file=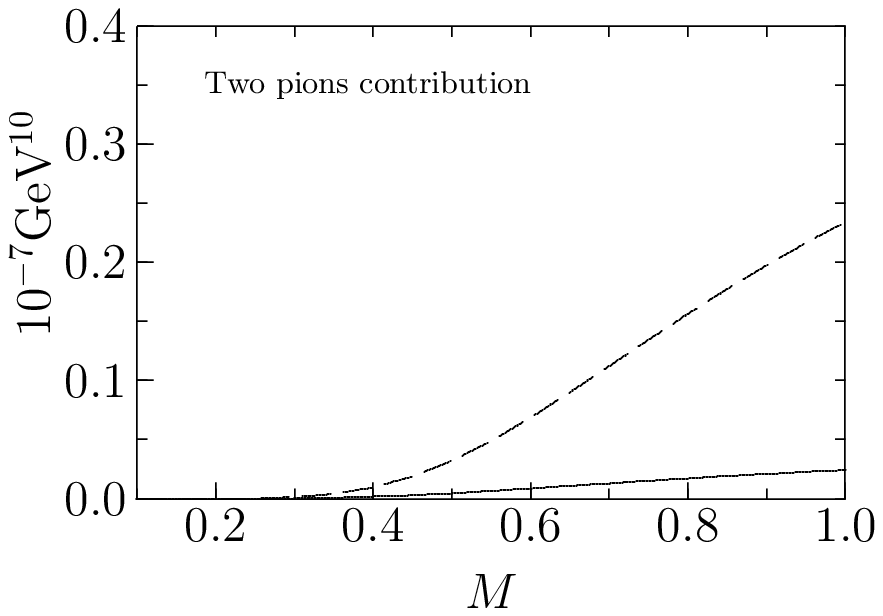,width=6cm,height=5cm}
\caption{The two pion contribution to the phenomenological side of
the SR with $\alpha~=~\beta=~1$. The solid (dashed) line contains
(does not contain) the effect of the form factor.}
\end{minipage}
\end{figure}
\begin{figure}[h]
\begin{minipage}[c]{8cm}
\vspace*{1cm}
\centerline{\epsfig{file=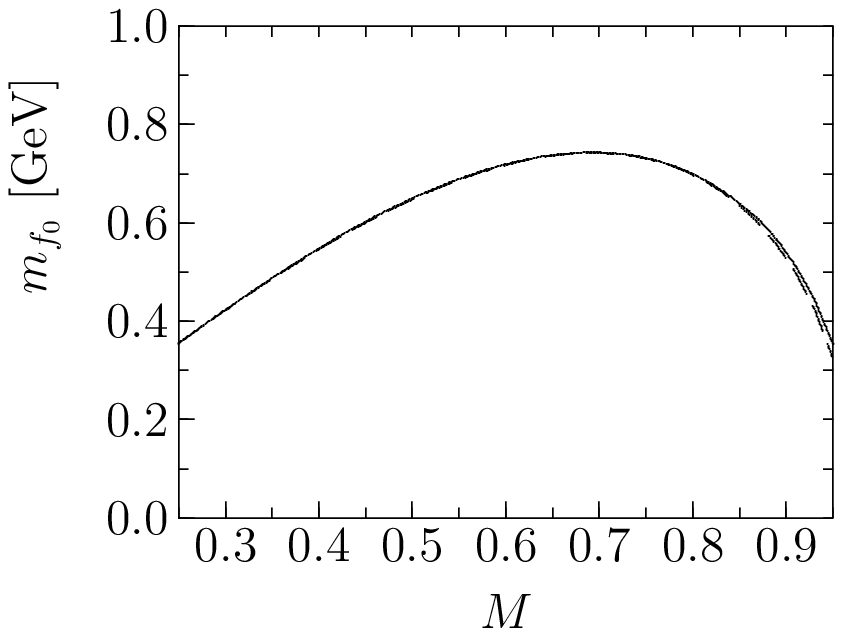,width=8cm,height=6cm,angle=0}}
\caption{The mass obtained from the SR for $\alpha~=~\beta~=~1 $
including (solid line) and not including (dashed line) the effect of
the form factor as a function of the Borel parameter. }
\end{minipage}
\hspace*{0.5cm}
\begin{minipage}[c]{8cm}
\vspace*{1.0cm}
\vskip 1.1cm
\centerline{\epsfig{file=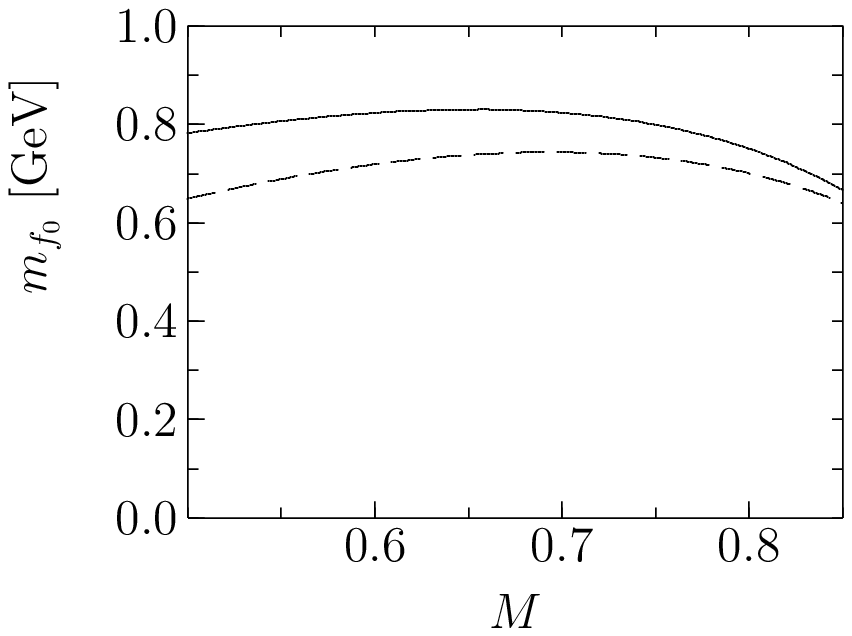,width=8cm,height=6cm,angle=0}}
\caption{The mass obtained from the SR for $\alpha=-\beta=1 $
including the effect of the form factor as a function of the Borel
parameter. The dashed line corresponds to the mass obtained from the
SR not including the two pion contribution.}
\end{minipage}
\end{figure}
In Figs. 2 and 3, the contribution from the OPE together with
direct instantons (the first two terms on the left hand side in
Eq.(\ref{SR})) and the contribution from two pion state (dashed
line) are shown as functions of the Borel mass $M$ for the maximum
chirality case of $\alpha=-\beta=1$. As result of the huge
contribution coming from the two pion state, the left hand side
(LHS) of the SR Eq.(\ref{SR}) becomes negative and therefore it is
impossible to obtain information about the resonance state with such
SR. The opposite situation we observe for the zero chirality case,
$\alpha=\beta=1$. In Figs. 4 and 5, the contributions from the OPE
with direct instantons and the two pions are shown for this case.
Now the contribution from the two pion state is very small and the
correspondent SR shows the signal for a possible bound state, as
shown in Fig.~6 by the dashed line.

However, we should point out that in the above calculation the
point--like local tetraquark--two pions vertex has been considered.
The simple way to include the effect of nonlocality into our
consideration is to introduce a form factor in the tetraquark--two
pions vertex. Such form factor for the scalar tetraquark $\sigma
(f_0(600))$ meson can be chosen of the monopole form
\begin{equation}
F_{f_0\pi\pi}(s^2)=\frac{1}{1+\langle r_\pi^2\rangle s^2/6}
\label{form}
\end{equation}
with the slope given by scalar radius of $\pi$ meson, $\langle
r_\pi^2\rangle\approx 0.75$ fm $^2$ \cite{yndurain}. Before fitting
the mass, it is necessary to fix the Borel window with the
pole contribution dominance \cite{Matheus07}.
These windows for the $\alpha=\beta$ and for the $\alpha=-\beta$
cases lie in $2m_\pi<M<0.93$ GeV and 0.5 GeV$<M<$0.85 GeV,
respectively. Our final result for mass of the tetraquark extracted
from the SR with the two pion contribution and the form factor
effect is presented in Figs.~6 and 7. It follows from the previous
figures that the effect from the form factor in the two pion state
is rather small for the interpolating current with $\alpha=\beta=1$.
Unfortunately, there is no good plateau of stability for the mass in
Fig.6. Therefore, it is rather difficult to extract the information
about $f_0(600)$ for this current. On the other hand, for the
interpolating current of $\alpha=-\beta=1$, the effect from the form
factor suppresses the two pion contribution and makes the SR have
physical meaning with a very good stability plateau presented in
Fig.7. The fitted mass of resonance for the $\alpha=-\beta $ case,
$m_{f_0}\approx 800$ MeV, is a little bit larger than the value
obtained recently from analysis of the Roy equations \cite{Roy}, but
still lies within the interval of mass for the $\sigma(f_0(600))$--meson,
$m_{f_0}(600)=400\div 1200$ MeV given by PDG \cite{PDG}. A
more exact definition of the properties of $f_0(600)$ meson might be
possible including the mixing with the ordinary $q\bar q$ state
or/and glueball. The additional shift of mass may arise also from
the contribution of multiinstanton configurations. In spite of the
fact that such contribution is expected to be suppressed by the
small packing fraction of instantons in QCD vacuum $
f=n_{eff}\pi^2\rho_c^4\approx 0.1 $~\cite{shuryak98}, it may lead to
the shift of the extracted mass of hadrons up to few ten percent in
the magnitude \cite{dorokhov}. In Fig.7 we also present the result
for $f_0(600)$ mass without the two pion contribution (dashed line).
It follows that the coupling to the two pion state leads to an increase
of the lightest tetraquark mass by a value of about 100 MeV.

\begin{figure}[h]
\centerline{\epsfig{file=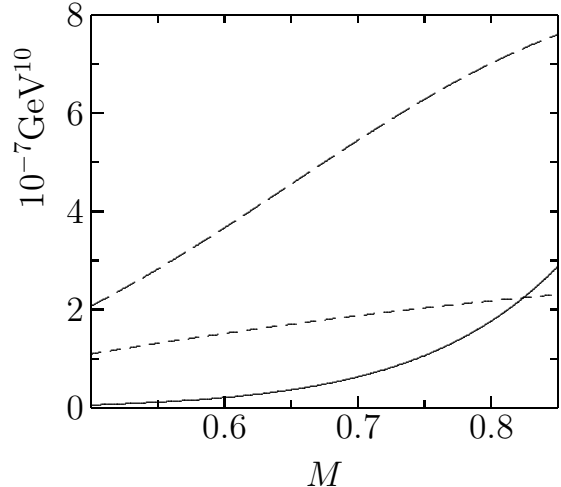,width=8cm,angle=0}} \caption{The
contributions to the SR for $\alpha=-\beta=1$ are shown: direct
instantons (long dashed line), two pion (short dashed line), and OPE
(solid line). }
\end{figure}
To clarify the origin of $f_0(600)$ state, in Fig.~8 we present
separately the contributions coming from standard OPE, direct
instantons and two pion state for $\alpha=-\beta $ case. It is
evident that direct instantons give the dominant contribution to the
SR. In spite of the fact that the two pion state contributes smaller
than direct instantons, its contribution is bigger than the
standard OPE within practically the full Borel window. It turns out
that we may treat in this case the $f_0(600)$ state as the
tetraquark bound state in the instanton field with a rather strong
coupling to the two pion state. The average size of instanton in the
QCD vacuum is rather small $\approx 0.3$ fm \cite{shuryak98},
therefore, the size of the lightest tetraquark should be also very
small. It would be interesting to find a possible experimental
signature for such small size for the $f_0(600)$. Finally, we should
emphasize that the crucial role of the direct instantons in the tetraquark
structure,  as shown above, does not allow us to agree with the
results of many tetraquark studies, where the direct instanton
contribution was not included.

\section{Conclusion}

In summary, the estimate of the two pion contribution to the QCD
sum rule for a light tetraquark by using the soft pion PCAC relations
has been obtained for the different tetraquark interpolating currents.
It has been demonstrated that such contribution depends crucially on
the chirality structure of the interpolating current. We show that
the SR for the interpolating current with maximum chirality provides
a very good stable plateau for the mass of the tetraquark when the
corresponding form factor in the tetraquark--two pion vertex is
introduced. The important role of the direct instantons in light
tetraquark dynamics has been demonstrated.

\section*{Acknowledgments}
We are happy to acknowledge useful discussions on various aspects
of this research with A.E. Dorokhov  and V. Vento. NK especially
grateful to Prof. D.--P. Min  for his kind hospitality at the
School of Physics and Astronomy of Seoul National University.
This work was supported by the research grant of the Chungbuk
National University in 2007 and by the Korea Research Foundation Grant
funded by the Korean Government (MOEHRD, Basic Research Promotion Fund)
(KRF-2007-313-C00177)(HJL). NK's work was supported in part by the Brain
Pool program of the Korea Research Foundation through KOFST grant 042T--1--1.

\end{document}